\documentclass[superscriptaddress,aip,showpacs,reprint]{revtex4-1}
\usepackage{graphicx}
\usepackage{amsfonts,amssymb,bm,ulem}
\usepackage[usenames]{color}
 \newcommand{\g}{pdf}
 
\begin{document} 

\title{Effect of hydrogen plasma treatment and oxygen deficiency on conducting properties of In$_2$O$_3$:Sn thin films} 
\author{V.~G.~Kytin}
\affiliation{Faculty of Physics, M. V. Lomonosov Moscow State University, 119991, Moscow, Russia}
\author{V.~A.~Kulbachinskii}
\affiliation{Faculty of Physics, M. V. Lomonosov Moscow State University, 119991, Moscow, Russia}
\author{O.~V.~Reukova}
\affiliation{Faculty of Physics, M. V. Lomonosov Moscow State University, 119991, Moscow, Russia}
\author{Y.~M.~Galperin}
\affiliation{Department of Physics, University of Oslo, PO Box 1048 Blindern, 0316 Oslo, Norway}
\affiliation{A.~F.~Ioffe Institute, Russian Academy of Sciences,  194021 Saint Petersburg, Russia}
\author{T.~H.~Johansen}
\affiliation{Department of Physics, University of Oslo, PO Box 1048 Blindern, 0316 Oslo, Norway}
\author{S.~Diplas}
\affiliation{Department of Chemistry and Centre for Materials Science and Nanotechnology, University of Oslo, Blindern, N-0315 Oslo, Norway}
\affiliation{SINTEF Materials and Chemistry, Forskingsveien 1, P.O. Box 124 Blindern, 0314 Oslo, Norway}
\author{A.~G.~Ulyashin}
\affiliation{SINTEF Materials and Chemistry, Forskingsveien 1, P.O. Box 124 Blindern, 0314 Oslo, Norway}
\begin{abstract}
Electrical conductivity, Hall effect and magnetoresistance of In$_2$O$_3$:Sn thin films deposited on a glass substrates at different temperatures and oxygen pressures, as  well as the films treated in a hydrogen plasma, have been investigated in the temperature range 1.5-300 K. The observed temperature dependences of resistivity were typical for metallic transport of electrons except temperature dependence of resistivity of the  In$_2$O$_3$:Sn film deposited in the oxygen deficient atmosphere. The electron concentration and mobility for the film deposited at 230$^\circ$C was larger than that for the film deposited nominally at room temperature. Short (5 minutes) treatment of the films in hydrogen plasma leads to the enhancement of electrical conductivity while longer (30 minute) treatment has the opposite effect. The electrical measurements were accompanied by AFM and SEM studies of structural properties, as well as by XPS analysis. Basic on structural and electrical measurements we conclude  the reduction process initiated by the hydrogen plasma provides essential modification of the ITO films surface. At the same time, electrical properties of the remaining (located beneath the surface layer)  parts of the ITO films remain mostly unchangeable. XPS analysis shows that grown in situ oxygen deficient ITO films have enhanced DOS between the Fermi level and the valence band edge. The extra localized states behave as acceptors leading to a compensation of $n$-type ITO.  That can explain  lower $n$-type conductivity in this material crossing over to a Mott-type hopping.
\end{abstract}

\maketitle

\section{Introduction}

Tin-doped indium oxide (ITO) is well known as having the best combination of large optical transparency in the visible range and high electrical conductivity comparing to other  transparent conducting oxides, such as SnO$_2$:F, ZnO:Al.~\cite{1} Therefore this material is widely used as transparent electrode in displays,~\cite{2} solar cells,~\cite{3} and as a heat filter~\cite{1} due to its high reflectivity in infrared range caused by the presence of conducting electrons. Large electrical resistivity of transparent electrodes limits the efficiency of relevant electronic devices. Therefore, decreasing the resistivity of In$_2$O$_3$:Sn films without noticeable drop of transparency remains an important practical task. Since the conducting properties of indium oxide films essentially depend on the preparation conditions~\cite{4} optimization of these conditions can pave a way to the resistivity reduction. A proper optimization requires understanding of the influence of different factors, like substrate temperature, atmosphere composition, etc. on electron transport in such films.

Depending on the processing conditions thin ITO films  ($\sim  \! \! 80$~nm) have different surface morphologies at nanoscale; they can be either  amorphous or nanocrystalline (or combination of both) in depth.~\cite{5,5a,5b,5c,5d}  In some cases the films consist of nanograins separated by insulating barriers.
As a result, electrical properties of ITO layers processed, e.g., by wet chemical methods can significantly differ.~\cite{6a} Therefore, it is important to analyze  the transport properties of the ITO layers prepared by  physical methods, which are commonly used for the device processing. The most popular method is  magnetron sputtering of ITO layers from metallic InSn or sintered ITO targets.  It is well known that structural properties of ITO layers can be essentially modified by hydrogen plasma treatments.  The plasma treatment results in formation of nanostructured films due to chemical reduction of indium and tin oxides to their corresponding metallic elements.~\cite{6b,6c,6d,6e,6f,6g}  Since the reduction process under special conditions can lead to formation of  nanostructured film~\cite{6g} one can expect the electrical properties, especially at low temperatures, will significantly differ from those of the 
 stoichiometric ITO layers.

 Low-temperature electrical properties of both as deposited and heat treated ITO layers were investigated in a number of works.~\cite{6aa,6ab,6ac,6ad} In particular, Hall measurements~\cite{6aa} of highly conductive ITO films prepared by DC magnetron sputtering on glass substrates in the temperature domain 6-300 K revealed metallic behavior.  Systematic measurements of resistivity~\cite{6ab} between 1.8 and 300 K of the transparent  films deposited using the standard RF sputtering technique also show metallic, free-electron-like conductance.  Amorphous and polycrystalline ITO films prepared by electron-beam evaporation were studied in Ref.~\onlinecite{6ac}.  The amorphous films behaved as  semiconductors, while polycrystalline samples behaved as metals.  Their magnetoconductivities are positive at low temperatures and can be described by the theory of  three-dimensional weak localization. Porous thin films with  In$_2$O$_3$:Sn-nanoparticles  prepared by a wet chemical technique were characterized by measurements of their temperature-dependent electrical resistivity.  The results were interpreted as  fluctuation-induced tunneling between micrometer-size clusters of internally connected ITO particles.~\cite{6ad} 

Electrical resistance and thermopower of a set of RF sputtered and annealed ITO films were measured from 300 K down to liquid-helium temperatures in Ref.~\onlinecite{6ae}. According to these measurements, between 150 and 300 K  the film resistance can be interpreted along the classical theory of metallic conductance. At lower temperatures, quantum contributions to the conductivity become important.  They can be interpreted as manifestation of weak-localization effects and electron-electron interaction in two-dimensional systems.~\cite{6ae}   Thus, electrical properties of ITO layers can be affected by the level of disorder or the presence of nanograins.

The present work is aimed to clarify  the influence of hydrogen plasma and oxygen deficiency on the low temperature  conducting properties of  In$_2$O$_3$:Sn films deposited by magnetron sputtering on glass substrates. We are not aware of previous similar studies. We report here the results of the investigation of electrical resistivity, Hall effect and magnetoresistance in the temperature range from 1.3 to 293 K for tin doped indium oxide films deposited and treated at different conditions.  

\section{Experimental}

Thin (70-80 nm) ITO layers were deposited on Corning glass by DC magnetron sputtering  at 230$^\circ$C (ITO-230) and at nominally room temperature (ITO-RT). In the latter case, some unintentional heating up to $\sim \! \! 60^\circ$C occurred during the deposition process. An ITO sintered target with In$_2$O$_3$ and SnO$_2$ in a weight proportion of 9:1 was used. The base pressure in the sputter system was about 10$^{-5}$ Torr. The total pressure of sputtering gas mixture was adjusted at $3\cdot 10^{-3}$ Torr during the film preparation. The Ar flow rate and the DC plasma power were kept constant of 38 standard cubic centimeters per minute (sccm) and 100 W respectively at all deposition temperatures.  
Oxygen deficient ITO layers were processes by two methods: (i) sputtering of an In/Sn target in a reduced oxygen ambient, (ii) hydrogen plasma treatment, which provides reduction conditions for any oxide. 

To produce an oxygen deficient ITO layer we used a metallic In/Sn target for a single magnetron powered in DC mode in combination with the Plasma Emission Monitor (PEM) control.~\cite{6af} The value of the PEM set point was reduced twice compared to that for the processing of the ITO layers with highest transmittance and lowest resistivity. Optical properties of this oxygen deficient ITO layer have been reported in Ref.~\onlinecite{5b} where such  films were investigated to study the effect of oxygen deficiency on the electron transport. 
To investigate the influence of atomic hydrogen on the properties of the ITO-230 layers we performed hydrogenation of the samples in a PECVD setup using a 13.56 MHz plasma generator. The RF plasma density during the hydrogenation process was 0.04 W/cm$^2$, while the ignition of plasma was done at higher densities (above 0.1 W/cm$^2$). At the stage of the plasma ignition the samples were removed from the plasma area and then reintroduced after stabilization of the discharge at the density of 0.04 W/cm$^2$ with a hydrogen gas flux of 200~sccm.

The surface morphology of the ITO layers was analyzed by AFM using a Digital Instrument’s Nanoscope Dim 3100 microscope~\cite{Nanoscope} equipped with spreading resistance measurement electronics. The following four characteristic parameters for the analysis of the AFM measurements were used: (i) the Root Mean Square (RMS) Roughness ($R_q$), which gives the standard deviation within a given area; (ii) the Mean Roughness ($R_a$), which represents the arithmetic average of the deviations from the center plane; (iii) the difference in height between the highest and lowest points on the surface relative to the mean plane ($h_{\max}$); (iv)   the averaged differences of heights ($h_{\text{av}}$).~\cite{6af} The AFM measurements were performed in the tapping mode using commercial silicon tips MikroMasch NSC35/AlBS with typical tip curvature radius $< 10$ nm. In some cases the surface morphology of ITO layers was also examined by scanning electron microscopy (SEM).  Thickness measurements of as-grown and H$^+$ plasma treated ITO layers were carried out using a one-wavelength ($\lambda=633$ nm)   ellipsometry set up.  

Analysis of X-ray photoelectron spectra (XPS) was performed on a KRATOS AXIS ULTRA$^{\text{DLD}}$ X-ray photoelectron spectrometer using monochromatic Al K$\alpha$ radiation ($h\nu = 1486.6$ eV) at 15~kV and 10 mA.  The pass energies for the  survey and high resolution scan were 160 and 20 eV, respectively.
 Since the ITO films were deposited on insulating (glass) substrates the samples were mounted in a way ensuring an electrical contact with the sample holder and, therefore, with the spectrometer. Such a construction allows avoiding additional  charge neutralization. The pressure of $(3-6) \times 10^{-9}$  Torr in the chamber was maintained during the analysis.  Prior to the analysis the samples were slightly etched with Ar$^+$ (0.5 keV, 20~s) in order to remove C contamination and adsorbed O species.

Resistivity and Hall effect of the films have been measured by the conventional four probe method. For this purpose rectangle samples with typical dimensions $8$$\times$$2$~cm$^2$ were cut  from  substrates with deposited In$_2$O$_3$:Sn films. Electrical contacts were made from In-Sn alloy. For the measurements samples were mounted in the cryostat where the temperature was varied from 1.5 to 293 K. The Hall effect and magnetoresistance were measured in magnetic fields up to 0.7 T using an  electromagnet,  or  in the field up to 6 T  using a superconducting solenoid. 

\section{Results and discussion}
\subsection{Morphology of the ITO/glass layers}
AFM measurements of ITO-RT/glass structure show that the ITO layer is nanostructured having the following characteristics (nm):  
$$R_q = 0.7, \ \  R_a = 0.45, \  \ R_{\max} = 8.3, \  \    h_{\text{av}}  = 3.4. $$
At the same time, the ITO-230/glass structures consist of grains, which are completely flat on the nanoscale. More detailed AFM analysis of these samples were presented in Ref.~\onlinecite{5b}. 

According to previous TEM studies,~\cite{5} the ITO films deposited at room temperature on Si substrates are amorphous having some  crystalline inclusions. The films deposited at 230$^\circ$C from mixed oxide target consist of clearly seen crystalline grains with the size of 15-20 nm. 

\begin{figure}[h]
\centerline{
\includegraphics[width=8.3cm]{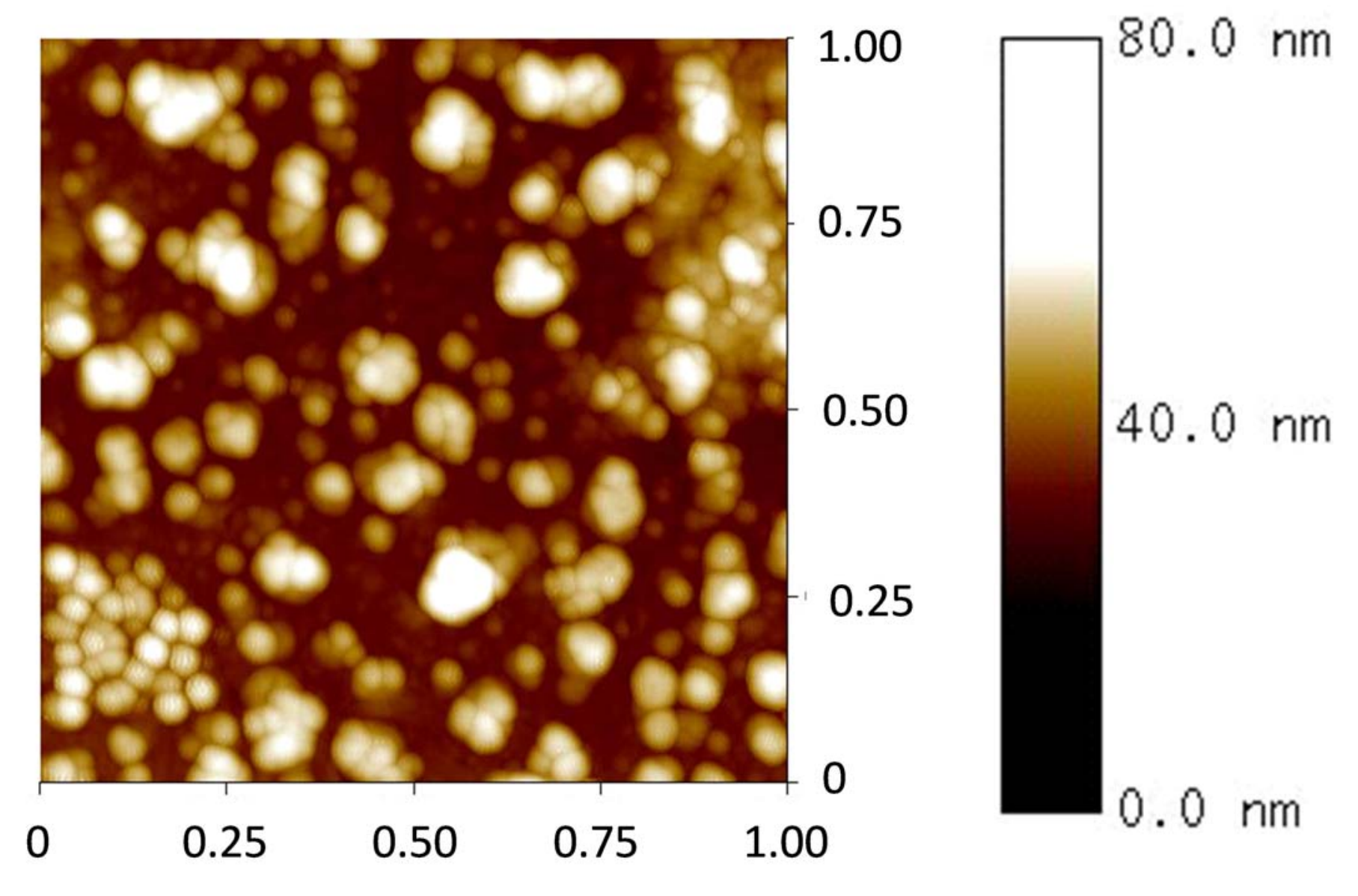}
}
\caption{AFM 2D image of ITO-230/glass structure after hydrogen plasma treatment at 230$^\circ$C for 5 min ($R_q = 6.5$ nm, $R_a = 5$ nm, $h_{\max} = 49$ nm, $h_{\text{av}} = 44$ nm). \label{fig01}}
\end{figure}

\begin{figure}[h]
\centerline{
\includegraphics[width=7cm]{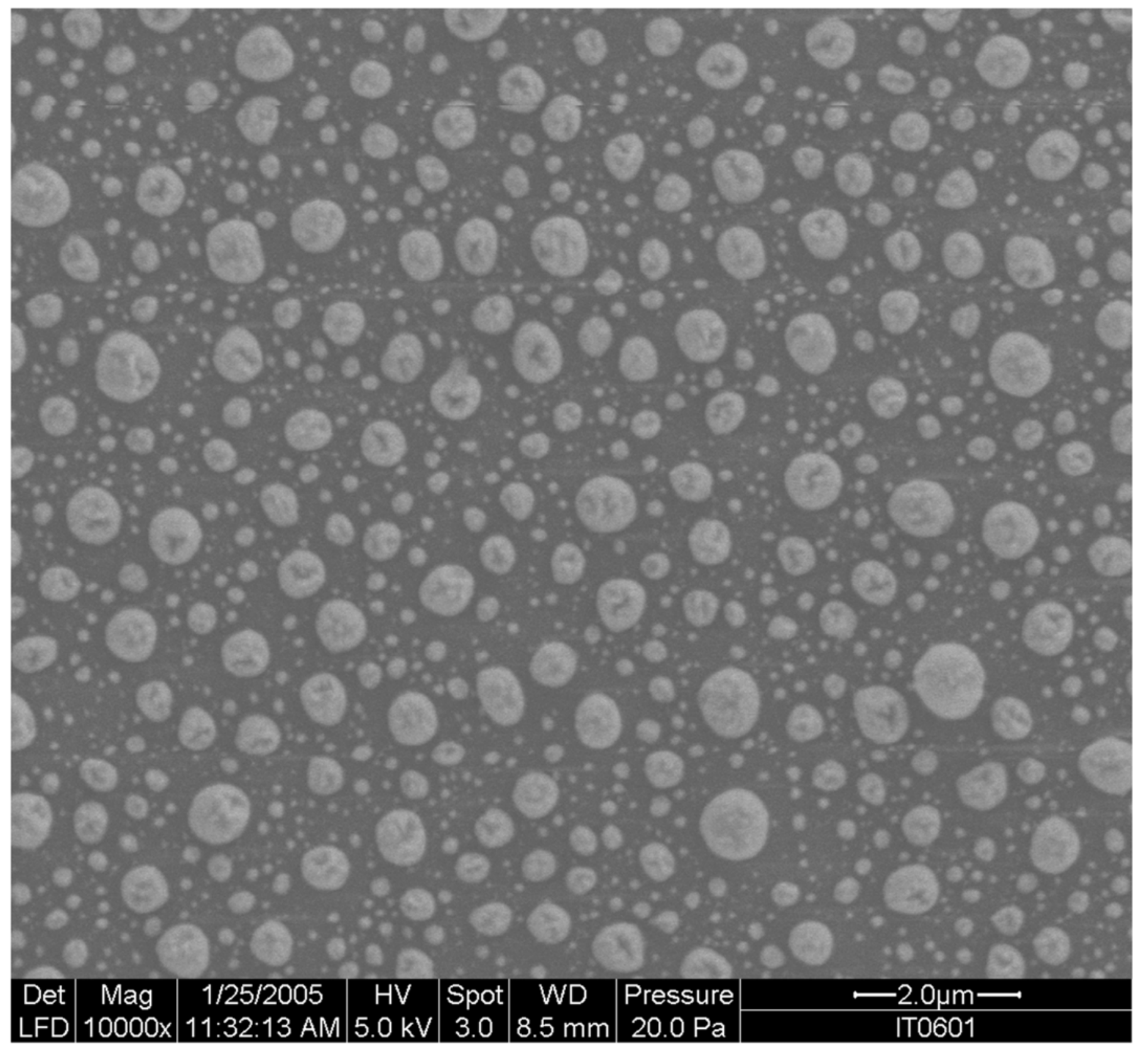}
}
\caption{SEM image of ITO-230/glass structure exposed to hydrogen plasma at 230$^\circ$C for 30 min \label{fig02}}
\end{figure}

Shown Fig.~\ref{fig01} are 2D images for ITO-RT/glass structure after hydrogen plasma treatment at 230$^\circ$C for 5 min. The AFM characteristics are given in the figure captions. As one can see, the plasma-treatment-induced chemical reduction leads to formation of a nanostructure. 
  Longer (30 min) plasma treatment  leads to  formation of metallic grains with the size of few microns (Fig.~\ref{fig02}) in agreement with Refs.~\onlinecite{6b,6c,6d,6e,6f,6g}. 
One can conclude that the reduction process initiated by the  hydrogen plasma provides essential modification of the  surface of the ITO films.  According to the results of the electrical measurements presented below, the modification of the surface layer does not significantly change the electrical properties of the part of the film located beneath the surface layer. 

\subsection{Hall effect: Electron concentration and mobility}

\begin{figure}[h]
\centerline{
\includegraphics[width=7cm]{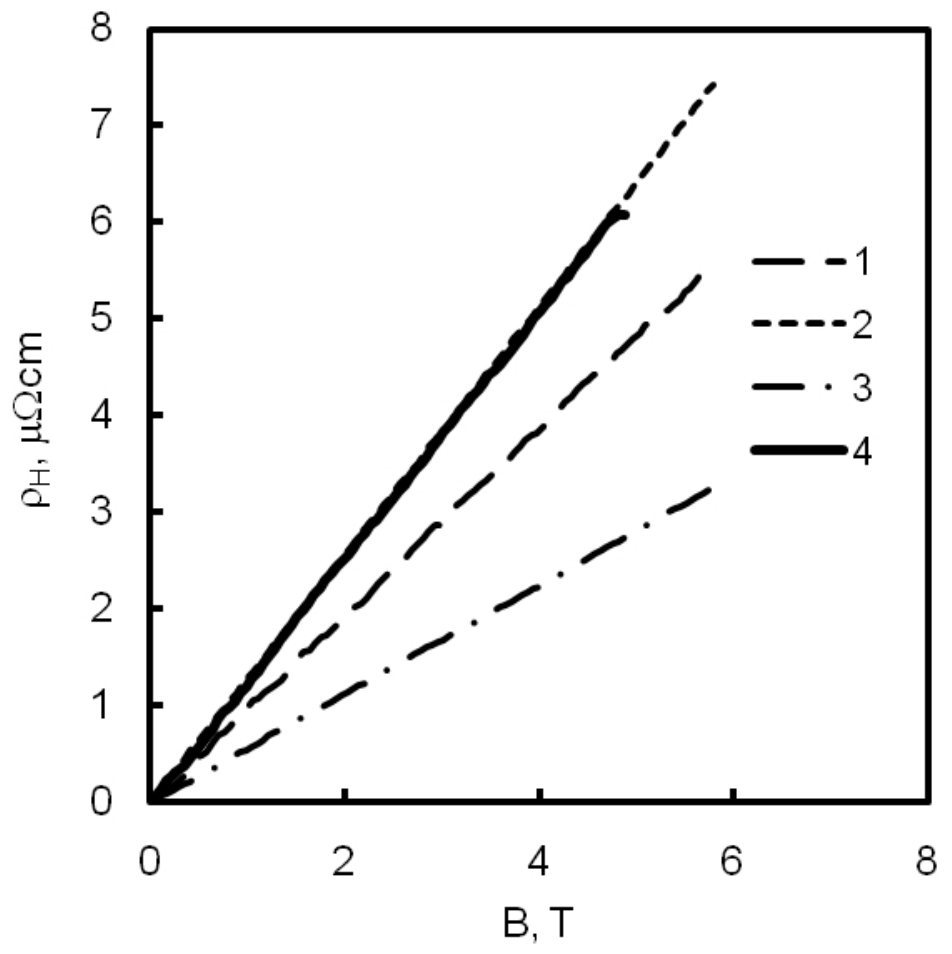}
}
\caption{ Hall resistivity, $\rho_H$, versus magnetic field, $B$ for the films 1-4 deposited from baked oxide target. $T= 4.2$ K. Numbers correspond to the Table 1. \label{fig1}}
\end{figure}

Shown in Fig.~\ref{fig1} are typical magnetic field dependences at 4.2 K of the Hall magnetoresistance of  films deposited from the baked In$_2$O$_3$:SnO$_2$ target.  These dependences are nearly linear in magnetic field  up to 6 T. 
They are used to calculate electron concentrations $n$, mobilities  $\mu$, Fermi energies $\varepsilon_F$, and electron mean free path $l$ using conventional expressions valid for Fermi statistics:
\begin{equation}
n=\frac{eB}{\rho_H d}, \  \mu=\frac{en}{\rho}, \  \varepsilon_F=\frac{(3\pi^2 n)^{2/3}\hbar^2}{2m^*} , \  l=\frac{\mu \sqrt{2m^*\varepsilon_F}}{e} . \label{eq1}
\end{equation}
Here   $\rho$ and $\rho_H$  are the  resistivity and the  Hall resistivity, respectively;  $B$ is the magnetic field, $d$ is the film thickness, $e$ is the elementary charge, $m^*$ is the electron effective mass. Calculated values are listed in Table~\ref{table1}.  The films deposited from a metal target under the oxygen deficiency conditions show nonlinear magnetic field dependences of the Hall resistivity at 4.2 K.
\begin{table}[h]
\centerline{
\begin{tabular}{|c|c|c|c|c|c|c|c|}\hline
Film&Thickness,  & $T_d$,& $t_p$,& $n\cdot 10^{-20}$, & $\mu$, & $\varepsilon_F,$ & $l$, \\ 
 \#&nm& $^\circ$C &  min & cm$^{-3}$& cm$^2$/Vs & eV & nm \\
\hline
1&80&\phantom{2}RT&\phantom{3}0&\phantom{3}6.4&11.6 & 0.77 & 2.0 \\
2&80&230 & \phantom{3}0&\phantom{3}4.9& 45\phantom{.3}& 0.64 & 7.2 \\
3&75&230 & \phantom{3}5 &11.7 &29.9 &1.15& 6.0\\
4&60& 230& 30 & \phantom{3}6.7 & 50.1 & 0.79& 6.7\\
\hline
\end{tabular}
}
\caption{Electron concentration $n$,  mobility $\mu$ , Fermi energy $\varepsilon_F$   and  mean free path $l$ in In$_2$O$_3$:Sn films deposited by magnetron sputtering from baked In$_2$O$_3$:SnO$_2$ target. $T_d$ is the deposition temperature while $t_p$ is the duration of H$^+$ plasma treatment. \label{table1}} 
\end{table}

For all  films deposited from a metal target  the estimated  mean free path is significantly shorter than the film thickness, and the estimated Fermi energy is much greater than the thermal energy not only at 4.2 K, but even at room temperature (0.026~eV). Therefore usage of Eq.~(\ref{eq1}) valid for a  three dimensional degenerate electron gas is justified.

\subsection{Temperature dependence of resistivity}

\begin{figure}[h]
\centerline{
\includegraphics[width=7cm]{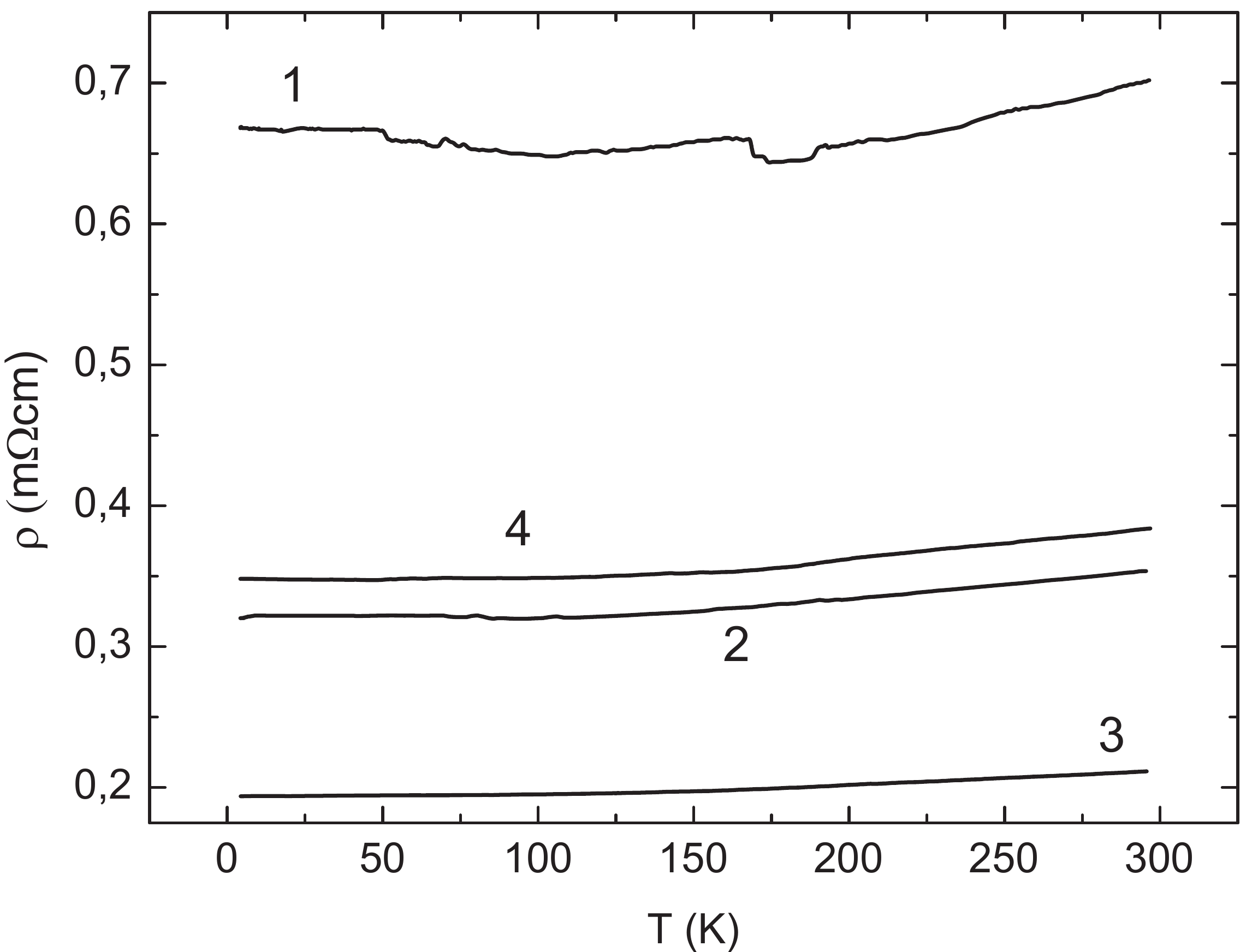}
}
\caption{Temperature dependence of resistivity of the films 1-4 deposited from a baked oxide target. Film numbers correspond to the Table~\ref{table1}. \label{fig2}}
\end{figure}

All the films deposited from the baked oxide In$_2$O$_3$:SnO$_2$ target show metal type temperature dependence of resistivity in the range 77-293 K, i.e., their resistivity increases with temperature. 
Typical temperature dependencies of the resistivity are shown in Fig.~\ref{fig2}. Temperature variation of the resistivity of the films deposited from an oxide target does not exceed 10\%.  For the films deposited at RT, resistivity  exhibits a minimum near 100 K. The  resistivity  of the films deposited at 230$^\circ$C saturates below 100 K.  
At all temperatures, the resistivity of the films deposited at 230$^\circ$C is smaller than the resistivity of the films deposited at RT. According to the Table~\ref{table1} this is mainly due to the higher electron mobility. This is consistent with previously reported data~\cite{6} as well with the fact that  the films deposited at 230$^\circ$C are more crystalline.~\cite{7}

Treatment of the films in hydrogen plasma during 5 min lowers the resistivity of the films and increases electron concentration. The  increase of electron concentration can be explained by the diffusion of protons inside the films. The protons can act as donors supplying additional electrons.~\cite{9} Longer (30 min) treatment of the In$_2$O$_3$:Sn films in  hydrogen plasma reduces the electron concentration  to the value close to the  concentration in untreated films. The origin of this effect is unclear. 

\begin{figure}[h]
\centerline{
\includegraphics[width=8.3cm]{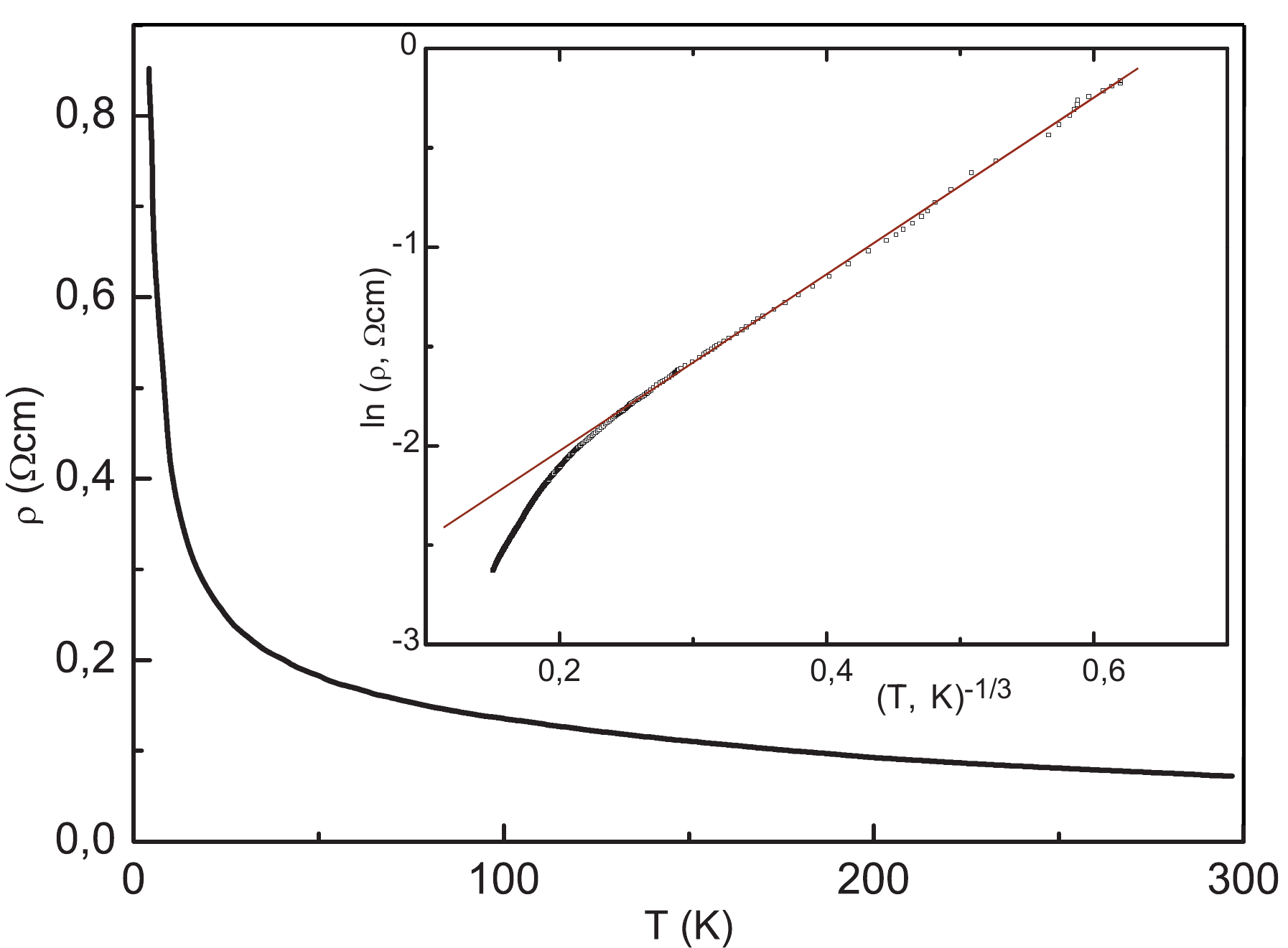}
}
\caption{Temperature dependence of resistivity of the film deposited from metal target in the oxygen deficient atmosphere. Inset: Logarithm of the resistivity versus $T^{-1/3}$. Symbols are experimental points. Line is the best fit of the experimental points  using Eq.~(\ref{eq2}) in the temperature domain 4.2-125 K. \label{fig3}}
\end{figure}

The resistivity of films deposited from a metal target in an atmosphere with reduced oxygen content increases while lowering the temperature (Fig.~\ref{fig3}).  
The low-temperature part of this dependence is consistent with the variable range hopping mechanism.~\cite{10}  However, it is difficult to decide whether the hopping is two (2D) or three dimensional (3D).
Below 80 K the experimental temperature dependence of the resistivity reasonably agrees  with Mott's law for 2D systems: 
\begin{equation} \label{eq2}
\rho =\rho_0\exp\left[(T_0/T)^{1/3}\right]\, .
\end{equation}
The best fit of the experimental  dependence to Eq.~(\ref{eq2}) yields  $T_0=87$ K. This quantity can be used to estimate the lower limit for the localization length, $r_{\mathrm{loc}}$, using the expression:~\cite{10}
\begin{equation} \label{eq3}
  r_{\mathrm{loc}}=\sqrt{13.8/k_BT_0 g_{\mathrm{loc}}(\varepsilon_F)}  \, ,          
\end{equation}
where $g_{\mathrm{loc}}(\varepsilon_F)$ is the (2D) density of localized states at the Fermi energy.  As an upper limit for this quantity we take the 2D density of extended states,
\begin{equation} \label{eq4}
g(\varepsilon_F)=m^*/\pi \hbar^2
\end{equation}
where $m^*=0.35m_0$ is the electron effective mass in In$_2$O$_3$, while $m_0$ is the free electron mass.~~\cite{6}  Substitution of Eq.~(\ref{eq4}) in  Eq.~(\ref{eq3})    gives $r_{\mathrm{loc}}=35$ nm. This is smaller, but comparable with the film thickness 80 nm. Below 20 K the temperature dependence of the conductivity can  also be  fitted by an expression derived for VRH in the presence of Coulomb gap near the Fermi energy:~\cite{10}
\begin{equation} \label{eq5}
\rho =\rho_0\exp\left[(T_1/T)^{1/2}\right]\, , \quad k_B T_1=e^2/4\pi \varepsilon_0 \varepsilon_d r_{\mathrm{loc}}\, .
\end{equation}
In this expression $\varepsilon_0$  is the vacuum dielectric constant while  $\varepsilon_d=8.9$ is the relative dielectric constant of indium oxide.~\cite{6}  Best fit of the experimental temperature dependence of the resistivity by Eq.~(\ref{eq5}) is provided by $T_1=20$ K. This corresponds to the estimate  $r_{\mathrm{loc}}=95$ nm, which is more consistent with a  2D character of hopping conduction. 
Both estimates of the localization length are essentially larger than the effective Bohr radius, $a^*=1.3$ nm in In$_2$O$_3$.~\cite{8} This points to the fact that localized states responsible for hopping conductivity are formed by donor clusters or areas with enhanced donor density.

\subsection{Magnetoresistance}

The variation of the resistivity caused by magnetic field for the films deposited at 230$^\circ$C and treated in hydrogen plasma (films 3 and 4)  is less than 0.01\%, i.e., on the order of the  noise level.  
\begin{figure}[h]
\centerline{
\includegraphics[width=7cm]{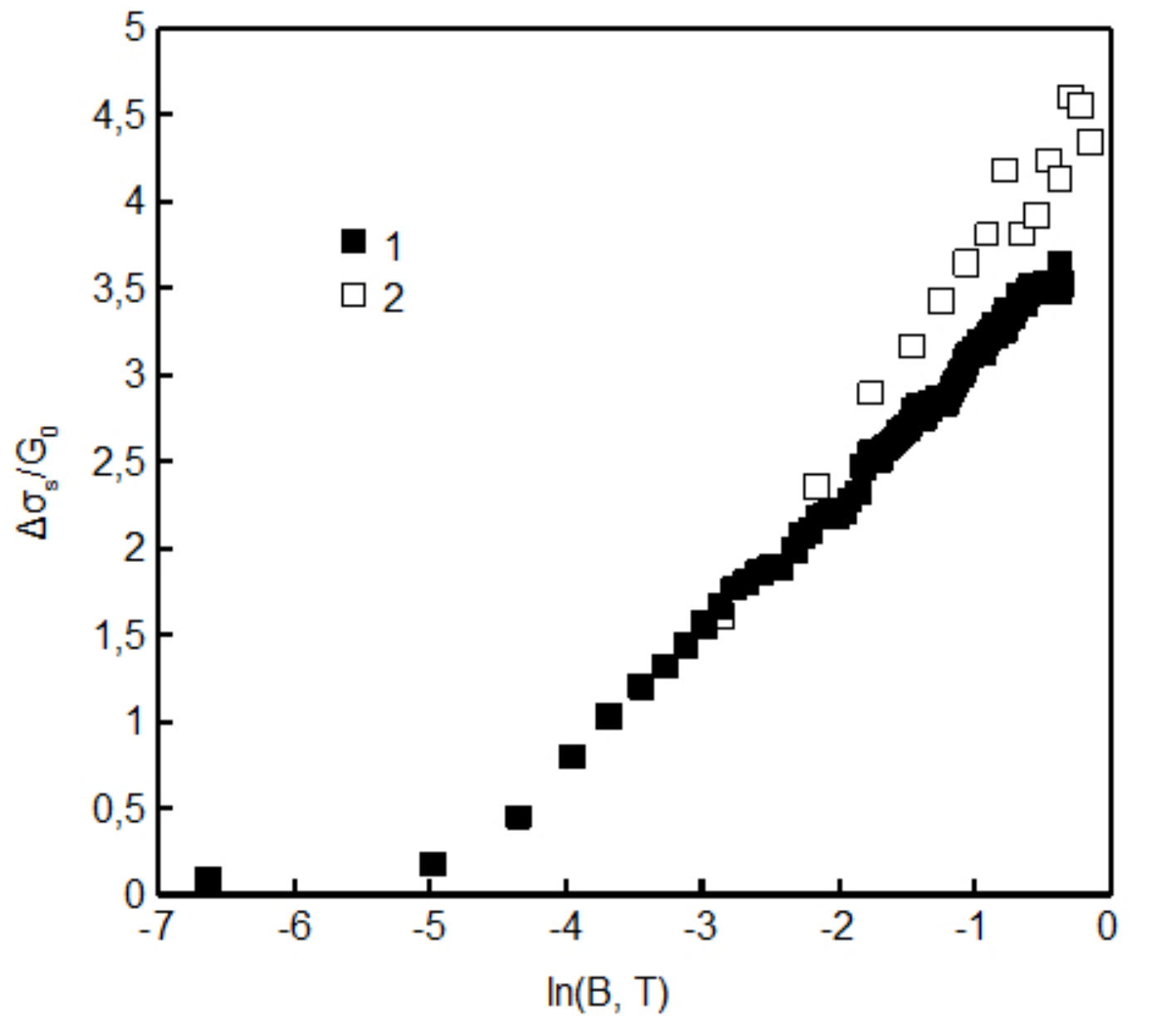}
}
\caption{Magnetoconductivity of the films 1 and 2  deposited from baked oxide target. $T=  4.2$ K. Film numbers correspond to the Table~\ref{table1}. 
$\Delta \sigma_s \equiv \left[ \rho^{-1}(B)- \rho^{-1}(0) \right]d$, $G_0 = 1.235\times 10^{-5}$  $\Omega^{-1}$.\label{fig4}}
\end{figure}

The magnetoresistance of the films 1 and 2 deposited at 25$^\circ$C and 230$^\circ$C is negative at low temperatures. 
Shown in Fig.~\ref{fig4} is the variation of the conductivity  at 4.2 K as a function of magnetic field for the films 1 and 2.  At $B \ge 0.1$ T,  magnetoconductivity is roughly proportional to $\ln B$, the proportionality factor being close to 1   if the conductivity is calculated per square unit of the film and expressed in the units of $G_0=1.233 \times 10^{-5}$ $\Omega^{-1}$. The negative magnetoresistance can be understood in the framework of  the theory of weak localization. An approximate expression for the magnetoconductivity of a quasi-2D electron gas reads as:~\cite{11}
\begin{equation}\label{eq6}
\Delta \sigma_s/G_0=\ln (4eD\tau_\varphi B/\hbar)-1.96 \, .
\end{equation}
Here  $D$ is the electron diffusion constant, and $\tau_\varphi$ is the phase relaxation time of the electron wave function. The effective dimensionality of the
weak localization problem is determined by comparison of the film thickness, $d$,  with the diffusion length of an electron during the phase relaxation time,   $L_\varphi =\sqrt{D\tau_\varphi}$.  Fitting of the experimental magnetoresistance with Eq.~(\ref{eq6}) yields $ L_\varphi=230$~nm for the film deposited at 25$^\circ$C and 360 nm for the film deposited at 230$^\circ$C. Both values of  $L_\varphi$ is essentially larger than the film thickness. Thus the interpretation of the magnetoresistance as quasi-2D is self-consistent.  

\begin{figure}[h]
\centerline{
\includegraphics[width=7cm]{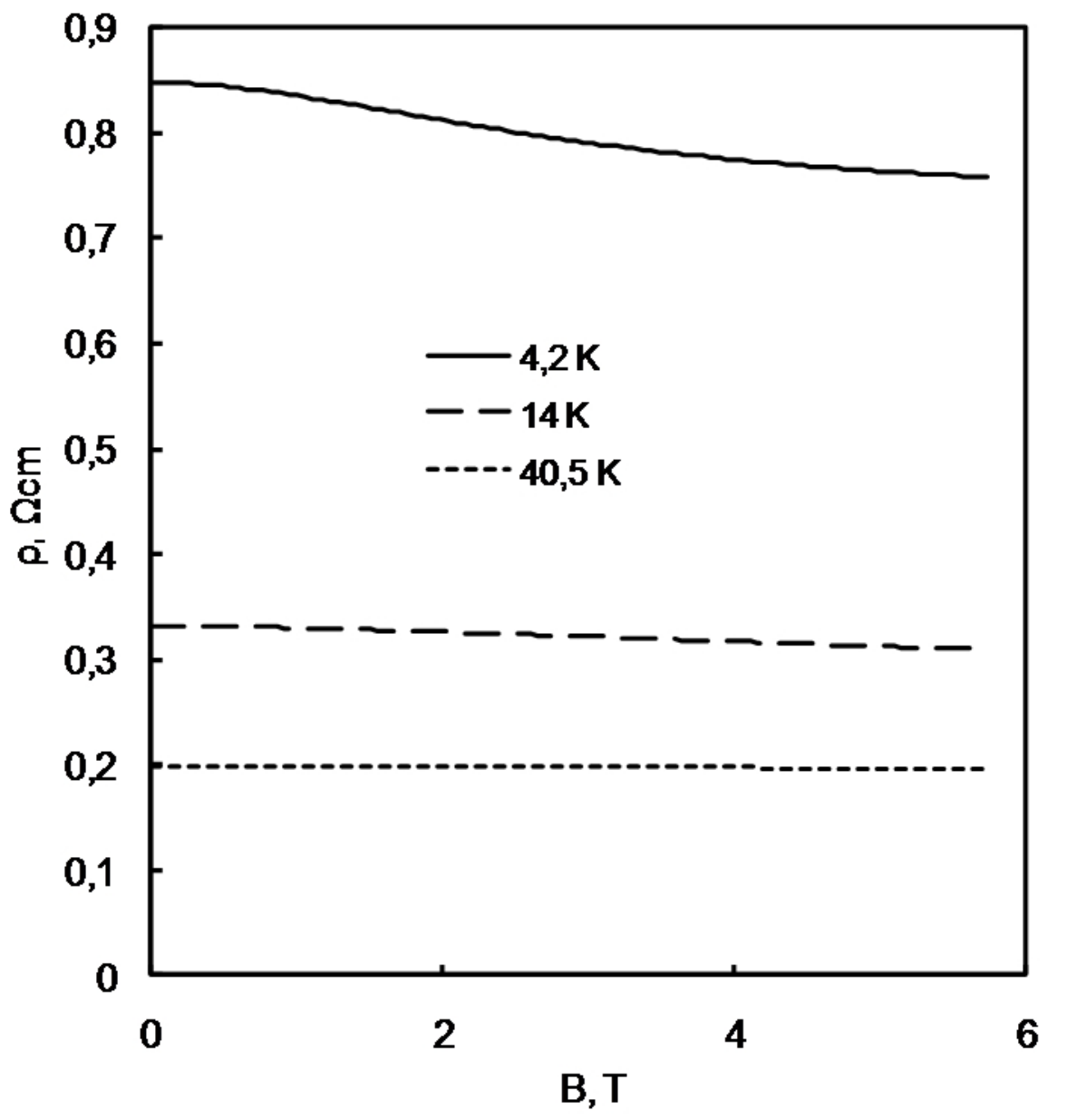}
}
\caption{Magnetoresistance of the film deposited from metal target in oxygen deficient atmosphere measured at 4.2 K, 14 K and 40.5 K as indicated.\label{fig5}}
\end{figure}

The magnetoresistance of the film deposited from a metal target in the oxygen deficient atmosphere is shown in Fig.~\ref{fig5}. 
The magnetoresistance is negative, and its magnitude  decreases when the temperature increases. A similar behavior was earlier observed in amorphous indium oxide.~\cite{12}  Our In$_2$O$_3$:Sn film deposited from a metal target in the oxygen deficient atmosphere seems to be similar to the amorphous  films studied previously.~\cite{12} The negative magnetoresistance (NM) in this film persists up to  B=6 T. In such magnetic field the quantum magnetic length, $l_B=\sqrt{\hbar/eB}$,  is equal to 10 nm. This value is essentially smaller than the estimated  localization length obtained from the temperature dependence of the resistivity. The origin of the observed NM remains unclear. Most probable explanation would be either a reduction of the gap near the Fermi energy or an increase of density of states near the Fermi energy at the higher energy induced by magnetic field. 

\subsection{XPS analysis}
\begin{figure}[h]
\centerline{
\includegraphics[width=8.3cm]{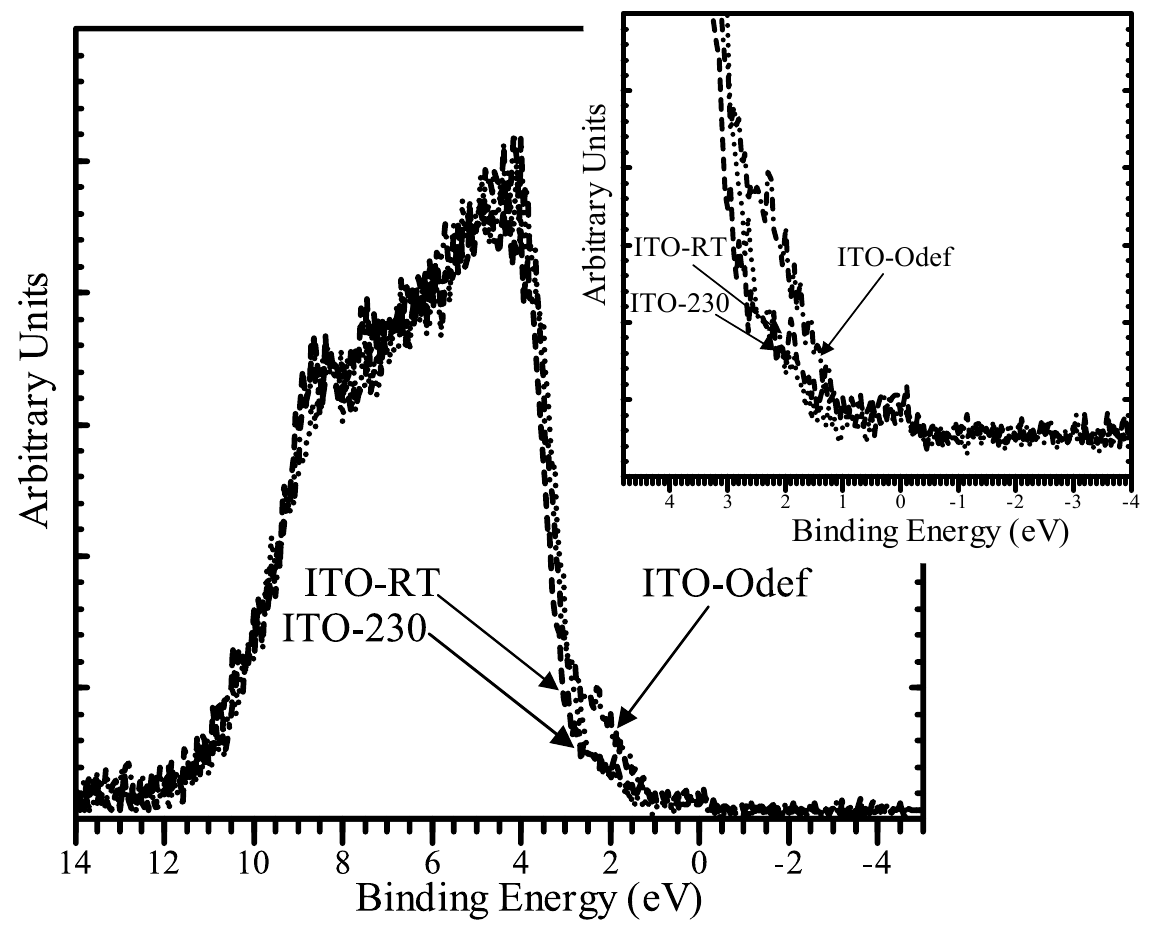}
}
\caption{Valence band XPS spectra of ITO films deposited at different conditions. Enhanced DOS close to the Fermi level is clearly seen from the inset where this region is zoomed in. \label{fig_XPS_1}}
\end{figure}

Figure~\ref{fig_XPS_1}  shows the valence band spectra for the ITO-RT, ITO-230 and oxygen-deficient ITO/glass structures.  One can see that the oxygen deficient film has larger density of states (DOS) between the Fermi level ($E_F$, which is set at the origin for the energy axis) and the valence band edge (VB).
These localized states serve as a sink for the electrons explaining lower $n$-type conductivity in the oxygen deficient material. In other words, such localized states near the VB can be considered as acceptor states, which cause compensation of the originally $n$-type ITO. 
It is worth noting that formation of such acceptor levels indicates that undoped oxygen deficient indium oxide can be a $p$-type material.  Detailed studies of such materials fall out of the scope of the present work.  We will present relevant results elsewhere.
 
\begin{figure}[h]
\centerline{
\includegraphics[width=7cm]{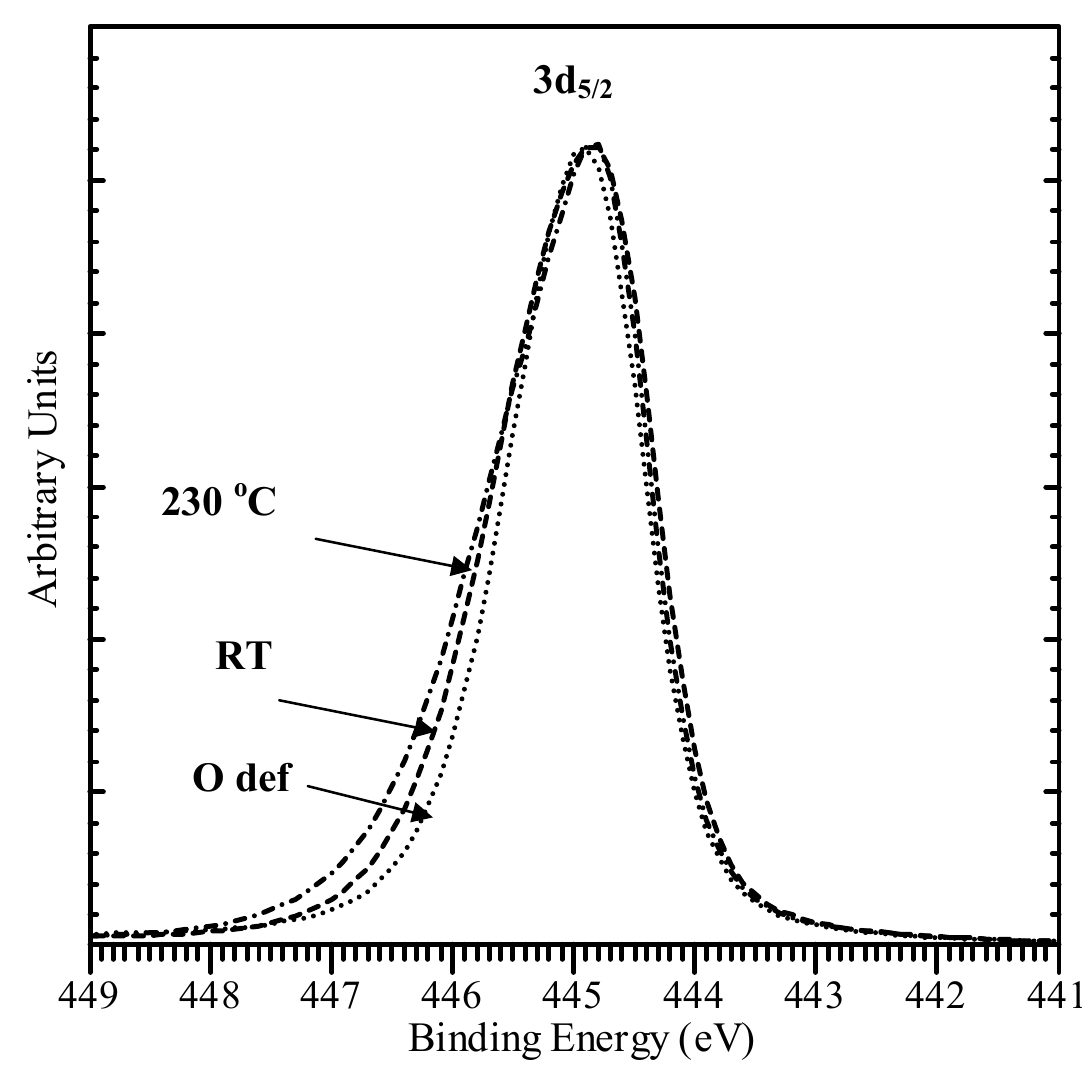}
}
\caption{High resolution In 3d XPS spectra of ITO fims deposited at different conditions   \label{fig_XPS_2}}
\end{figure}

\begin{figure}[h]
\centerline{
\includegraphics[width=7cm]{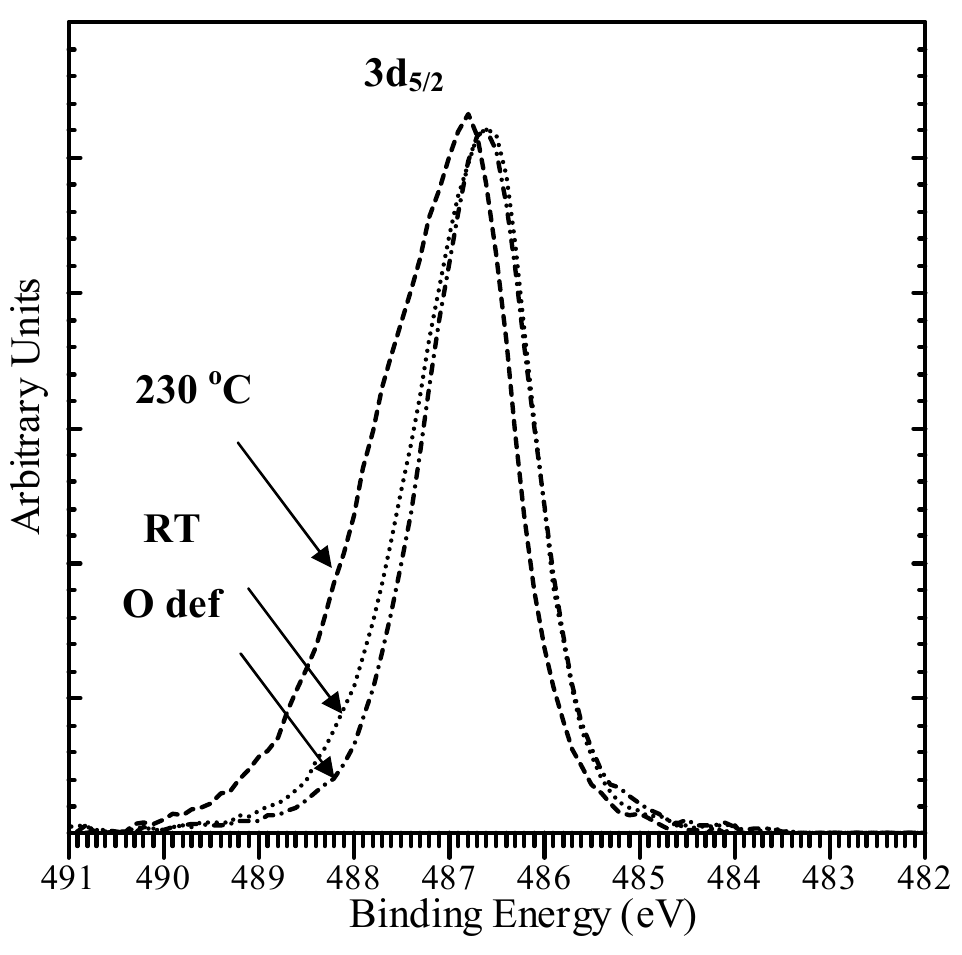}
}
\caption{High resolution Sn 3d XPS spectra of ITO fims deposited at different conditions   \label{fig_XPS_3}}
\end{figure}

Shown in Fig.~\ref{fig_XPS_2} are high resolution In 3d spectra for ITO films deposited as different conditions.
 According to this figure,  the asymmetry of the In 3d spectra is increased along the line ITO-Odef$<$ITO-RT$<$ITO-230. The higher peak asymmetry corresponds to a more conductive film.~\cite{13,14} A similar, even more pronounced trend can be seen from the high resolution Sn 3d spectra of ITO films deposited at different conditions, see Fig.~\ref{fig_XPS_3}. 

XPS-based results for the composition of the ITO films deposited on glass substrates at different conditions are given in Table~\ref{table2}, the quantification being based on the peak areas.
\begin{table}[h]
\centerline{
\begin{tabular}{|c|c|c|c|c|}\hline
ITO film & In 3d, \text{at.}\% & O 1s,  \text{at.}\% &Sn 3d,  \text{at.}\% & $\frac{\text{O}}{\text{In+Sn}}$\\ \hline
O deficient & 47.93 &48.03&4.04&0.92 \\
RT & 47.41 &48.94 & 3.65&0.96 \\
230$^\circ$C & 47.66 & 48.83 &3.50 & 0.95 \\
\hline
\end{tabular}
}
\caption{Composition of ITO films deposited at different conditions. Quantification is based on areas of the high resolution In 3d, O 1s and Sn 3d spectra. The last column to the right illustrates the O/(In+Sn) ratios
 \label{table2}} 
\end{table}

\section{Summary}

We have studied the effects of substrate temperature, treatment in a hydrogen plasma and oxygen deficiency on the conducting properties of In$_2$O$_3$:Sn films. The increase of substrate temperature from RT to 230$^\circ$C leads to essential increase of the electron mobility and film conductivity in agreement with earlier reported results. Short (5 min) treatment in a hydrogen plasma increases electron concentration, as well as the film conductivity. Longer (30 min) treatment in hydrogen plasma reduces the  electron concentration. 

It is established that the reduction process initiated by the hydrogen plasma provides essential modification of the ITO films surface. At the same time, electrical properties of the remaining (located beneath the surface layer)  parts of the ITO films remain mostly unchangeable. 

  Grown in situ oxygen deficient ITO films have enhanced DOS between the Fermi level and the valence band edge. Formation of these localized states explains lower $n$-type conductivity in this material.  The extra localized states can be considered as acceptors leading to a compensation of the $n$-type ITO. 
The oxygen deficient  ITO films have reduced electron conductivity, which crosses over from the metallic mechanism to the hopping one. 
{}From this point of view such samples, which are $n$-type degenerate semiconductors, shows the behavior similar to that observed in several conventional Si-based highly doped semiconductor structures, such as Si:P, Si:Sb, Si:Se, and Si:S.\cite{ad1,ad2,ad3} In particular, in these structures a low-temperature metal-to-insulator transition (MIT) was observed. In the insulator phase of these samples, the activation energy can be essentially affected by partial disordering.~\cite{ad3}
On the contrary, properly prepared ITO structures - both polycrystalline (ITO-230) and amorphous (ITO-RT) - do not show any trace of the MIT.  Therefore one concludes that the structural disorder weakly contributes to the electron scattering, which agrees with recent simulations\cite{ad4} for pure and tin-doped indium oxide.  The metallic character of the conductance indicates that in  stoichiometric ITO structures the Fermi level is located in the conduction band at any temperature within the studied temperature domain. The transition to the insulating behavior occurs only in non-stoichiometric materials, such as those prepared under oxygen deficiency.

\acknowledgments{
Financial  support by the Russian Foundation for Basic Research (project number 09-03-00942a) and by the  Norwegian Research Council (project number 185414/S30) is greatly acknowledged.}


\begin{thebibliography}{99}
\bibitem{1} I. Hamberg, C. G. Granqvist, J. Appl. Phys. \textbf{60}, R123 (1986).
\bibitem{2} C. G. Granqvist, A. Hultaker, Thin Solid Films \textbf{411}, 1, (2002).
\bibitem{3} H. Kobayashi, T. Ishida, K. Nakamura \textit{et al.}, J. Appl. Phys, \textbf{72}, 5288 (1992).
\bibitem{4} R. Tahar, T. Ban, Y. Ohya \textit{et al.}, J. Appl. Phys. \textbf{83},  2631 (1998).
\bibitem{5} S. Diplas, A. Ulyashin, K. Maknys \textit{et al.},  Thin Solid Films \textbf{515}, 8539 (2007). 
\bibitem{5a} K. Maknys, A. G. Ulyashin, H. Stiebig \textit{et al.}, Thin Solid Films \textbf{511-512}, 98 (2006).
\bibitem{5b} A. V. Mudryi, A. V. Ivaniukovich, A. G. Ulyashin, Thin Solid Films \textbf{515}, 6489 (2007).
\bibitem{5c}  M. L. Mottern, F. Tyhold, A. Ulyashin \textit{et al.}, Thin Solid Films \textbf{515}, 3918 (2007).
\bibitem{5d} G. Untila, A. Chebotareva, T. Kost \textit{et al.}, Thin Solid Films \textbf{515}, 8505  (2007)
\bibitem{6a} J. Ederth, P. Johnsson, G. A. Niklason \textit{et al.}, Phys. Rev. B \textbf{68}, 155410 (2003).
\bibitem{6b} O. Kuboi, Jap. J. Appl. Phys. \textbf{20}, L783 (1981).
\bibitem{6c} J. H. Thomas III,  Appl. Phys. Lett. \textbf{42},  794 (1983).
\bibitem{6d} S. Major, S. Kumar, M. Bhatnagar \textit{et al.}, Appl. Phys. Lett. \textbf{49},  394 (1986).
\bibitem{6e} R. Banerjee, S. Ray, N. Basu \textit{et al.}, J. Appl. Phys. \textbf{62},  912 (1987).
\bibitem{6f} J. Lan, J. Kanicki, Thin Solid Films \textbf{304}, 123 (1997)
\bibitem{6g}A. G. Ulyashin, K. Maknys, S. Diplas \textit{et al.}, Proc. WCPEC-4, IEEE \textbf{1}, 134 (2006).
\bibitem{6aa} N. Kikuchi, E. Kusano, H. Nanto \textit{et al.},Vacuum \textbf{59}, 492 (2000).
\bibitem{6ab}  Z. Q. Li, J. J. Lin, J. Appl. Phys. \textbf{96}, 5918 (2004).
\bibitem{6ac} X. D. Liu, E. Y. Jiang, and D. X. Zhang, J. Appl. Phys. \textbf{104}, 073711 (2008).
\bibitem{6ad} J. Ederth, P. Heszler, A. Hultaker \textit{et al.}, Thin Solid Films \textbf{445}, 199 (2003).
\bibitem{6ae} Bo-Tsung Lin, Yi-Fu Chen, Juhn-jong Lin \textit{et al.}, Thin Solid Films \textbf{518}, 6997 (2010). 
\bibitem{6af} C. May, J. Strumpfel, Thin Solid Films \textbf{351}, 48 (1999).
\bibitem{Nanoscope} Digital Instrument’s Nanoscope Command Reference Manual V5.12 rb., (2001).
\bibitem{6} I. Hamberg, C. G. Granqvist, J. Appl. Phys., \textbf{60}, R123 (1986). 
\bibitem{7} T.H. Breivik, S. Diplas, A. G. Ulyashin, \textit{at al.}, Thin Solid Films \textbf{515}, 8479 (2007).
\bibitem{8} {\"U}.  {\"O}zg{\"u}r, Ya. I. Alivov,  C. Liu \textit{et al.}, Appl. Phys. \textbf{98}, 041301 (2005).
\bibitem{9} M. Mizuhashi, Thin Solid Films 70, \textbf{91}  (1980). 
\bibitem{10} B. I. Shklovskii, A. L. Efros, \textit{Electronic Properties of Doped Semiconductors}, Springer Verlag, Berlin, 1984.
\bibitem{11} T. A. Polyanskaya, Yu. V. Shmartsev, Sov. Phys. Semicond. \textbf{23}, 1 (1989).
\bibitem{12} V. F. Gantmakher, M. V. Golubkov, J. G. S. Lok, A.~K.~Geim, Zh. Eksp. Teor. Fiz. \textbf{109}, 1765 (1996)[Sov. Phys. JETP  \textbf{82}, 951 (1996)].
\bibitem{13} Y. Gassenbauer and A. Klein, J. Phys. Chem. B \textbf{110}, 4793 (2006).
\bibitem{14} Y. Gassenbauer and A. Klein, Solid State Ionics \textbf{173}, 141 (2004).
\bibitem{ad1} V. V. Abramov, V. A. Kulbachinskii, V. G. Kytin \textit{et al.},  Sov. Phys. Semicond. \textbf{26}, 495 (1992).
\bibitem{ad2}V. A. Kulbachinskii, V. G. Kytin, V. V. Abramov \textit{et al.}, Sov. Phys. Semicond. \textbf{26}, 1009 (1992).
\bibitem{ad3}V. V. Abramov, N. B. Brandt, V. A. Kulbachinskii, Sov. Phys. Semicond. \textbf{25}, 310 (1991).
\bibitem{ad4}J. Rosen, O. Warschkow, Phys. Rev B \textbf{80}, 115215 (2009).



\end{thebibliography}
\end{document}